%% file: main.tex
\documentclass[journal]{IEEEtran}
\usepackage{hhline} 
\usepackage{cite}
\usepackage{amssymb}
\usepackage{graphicx}
\usepackage{multirow}
\usepackage[table, dvipsnames]{xcolor}
\usepackage{color}
\usepackage{pifont}
\usepackage{array}
\usepackage{booktabs}

\definecolor{fancyBlue}{RGB}{119,158,203}

\newcommand{\etal}{$\;$\textit{et\,al.}} 

\newcommand*\rot{\rotatebox{90}}
\newcommand*\OK{\ding{51}}

\begin{document}

\title{Recent Advances in Local Energy Trading in the Smart Grid Based on Game--Theoretic Approaches}
\author{Matthias~Pilz, and Luluwah~Al-Fagih
	\thanks{M.Pilz and L.Al-Fagih are with Kingston University London.}
	\thanks{Faculty of Science, Engineering and Computing.}
	\thanks{Matthias.Pilz@kingston.ac.uk}
}

\markboth{}{Recent Advances in Local Energy Trading in the Smart Grid Based on Game--Theoretic Approaches}

% make the title area
\maketitle

\begin{abstract}
The global move towards efficient energy consumption and production has led to remarkable advancements in the design of the smart grid infrastructure. 
%However, the potential of the smart grid remains limited without the integration of renewable energy resources. 
Local energy trading is one way forward. It typically refers to the transfer of energy from an entity of the smart grid surplus energy to one with a deficit. 

In this paper, we present a detailed review of the recent advances in the application of game--theoretic methods to local energy trading scenarios.
%Two forms of trading are differentiated (direct and indirect trading) and the latest achievements within them are summarised. 
An extensive description of a complete game theory-based energy trading framework is presented. It includes a taxonomy of the methods and an introduction to the smart grid architecture with a focus on renewable energy generation and energy storage. Finally, we present a critical evaluation of the current shortcomings and identify areas for future research. 
\end{abstract}


\section{Introduction}
\label{sec:1_introduction}
\input{./sec/1_introduction.tex}

\section{Game Theory Taxonomy}
\label{sec:2_gtBasics}
\input{./sec/2_gtBasics.tex}

\section{Smart Grid Architecture}
\label{sec:3_etBasics}
\input{./sec/3_etBasics.tex}

\section{Game Theory for Energy Trading}
\label{sec:4_gtANDet}
\input{./sec/4_gtANDet.tex}

\section{Conclusion}
\label{sec:5_conclusion}
\input{./sec/5_conclusion.tex}

\section{Acknowledgements}
The authors would like to thank Jean-Christophe Nebel for useful comments and valuable discussions. This work was supported by the Doctoral Training Alliance (DTA) Energy.

\bibliographystyle{IEEEtran}
\bibliography{./full_bib}

%%%%\begin{IEEEbiography}%
%%%%[{\includegraphics[width=1in,height=1.25in,clip,keepaspectratio]{./fig/MatthiasPilz}}]{Matthias Pilz}
%%%%(S'17) was born in Frankfurt (Oder), Germany, in 1990. He received the B.Sc. and M.Sc. degrees in physics from the University of Jena, Germany, in 2012 and 2015, respectively.
%%%%In 2016, he joined the Department of Computer Science and Mathematics, Kingston University London, UK, where he is currently pursuing his Ph.D degree in smart energy. His current research interests include smart grid, game theory, renewable energy resources, machine learning, energy trading.
%%%%\end{IEEEbiography}
%%%%
%%%%\begin{IEEEbiography}
%%%%[{\includegraphics[width=1in,height=1.25in,clip,keepaspectratio]{./fig/L-Al-Fagih}}]{Luluwah Al-Fagih}
%%%%completed her Ph.D in financial mathematics at The University of Manchester, UK, in 2013. Prior to this, she received her B.Sc. and M.Sc in mathematics and financial mathematics from King's College London in 2006 and 2007 respectively. She has been a member of the School of Computer Science and Mathematics, Kingston University London since 2013. Her current research interests include applications of game theory and optimal stopping in energy, finance and cyber security.
%%%%\end{IEEEbiography}
%%%%
%%%%\vfill

\end{document}

%% file: sec/1_introduction.tex
\IEEEPARstart{I}{n} recent times, there have been substantial efforts to reduce energy consumption, with many countries committing to combat climate change by adopting the 2015 Paris Agreement~\cite{EuropeanCommission2016}. Restricting green house gas emissions is critical for limiting the global average temperature increase to the Paris target of 1.5 degrees Celsius above pre-industrial levels. The rise in electricity demand~\cite{ExxonMobil2016} along with efforts to limit global warming, pose a serious and complex problem. One part of the solution is the implementation of the smart power grid~\cite{Ipakchi2009}, i.e.~a technologically advanced, decentralised version of the current power grid. It includes two-way communication and energy transfer which in turn allow for innovation and efficiency gains. Furthermore, the idea of microgrids (MGs) comes into life. An MG describes a locally distributed collection of electricity sources and smart--users, e.g. a neighbourhood or a village that is itself connected to a bigger macrogrid.

Key elements of the smart grid are distributed energy storage and the integration of renewable energy (RE) resources. The persistent trend of price reduction for solar panels, i.e. Swanson's law~\cite{Swanson2006}, inspires to expect many more homes to be equipped with their own small-scale power plant in the future. To extract most value from them, there is a need for a thorough understanding of this technology. Fortunately, the introduction of smart meters means that a lot more data can be accessed to achieve this goal. Suitable models are needed to utilise this data and make energy usage more efficient. In particular local energy trading among or within multiple MGs yields promising perspectives, as we point out in this paper. 

The term \emph{energy trading} has been historically used to refer to the buying and selling of energy, e.g.~electricity and gas, in the wholesale markets such as the European Energy Exchange. It traditionally takes place between producers, retailers and traders as well as large industrial users. More recently, the use of the term energy trading was extended by Saad \etal~in \cite{Saad2011} to refer to the `local' transfer of energy between users within a smart grid. Since then, an array of literature has adopted this new definition (cf.~\cite{Ilic2012, RodrigoVerschae2016, Chis2016, Couillet2012} and~\cite{Bayram2014} for a comprehensive review of energy trading in the smart grid). In such a scenario, it is usually assumed that the demand could also be satisfied from the macrogrid, but at a higher cost. The task is to find optimal strategies for each entity such that a reductions of energy costs is realised. Additionally, this brings several technical advantages. Firstly, the inherent local usage of energy in such a system results in better power quality, i.e.~less voltage fluctuations, and even more directly in less line loss~\cite{Bayram2014}. Secondly, the system is more reliable, as it is safe from outages of the macrogrid. 

A commonly used approach to tackle the energy trading problem is based on single objective optimisation~\cite{Ramachandran2011,Matamoros2012,Chen2013}. It uses a centralised approach, where an independent controller is in charge of solving the optimisation problem. The solution is the amount of energy to be traded such that generation and transportation costs are minimised. This stands in contrast to the idea of a decentralised power grid. Indeed, a trading model needs to evaluate the behaviour of all participants and incorporates their individual preferences. As the actions of one influences all the others, Game Theory (GT) is a suitable method to choose. GT is a branch of mathematics that deals with the analysis of competitive situations, where the outcome of one participant does not only depend on their own strategy but also on the strategies of the others. It was first introduced for problems in economics~\cite{vonNeumann1944}, but is nowadays applied in many areas, such as biology~\cite{Hammerstein1994,Weibull1995,Dugatkin1998} and computer science~\cite{Shoham2008}. The game--theoretic approach internalises the decentralised structure of the smart grid. 

The use of game--theoretic methods for the smart grid in general can been seen in~\cite{Mohsenian-Rad2010, Belgana2015, Kumar2015, Wang2015, Nguyen2015, Mondal2015, Asimakopoulou2013, Ma2010, Yin2016, Karfopoulos2013, Marzband2016}, and has been extensively reviewed in \cite{Saad2012} (and~\cite{Eslahi-Kelorazi2016}). In this paper, we review the recent advances in the use of game--theoretic approaches for the smart grid specifically within the local energy trading framework as introduced in the recent literature.

%Energy trading refers to the transfer of energy from an entity that produces more energy than it needs to an entity with a deficit. 

This work is intended to be a standalone reference for the reader and thus provides an introduction to the game--theoretic principles used in the smart grid energy trading literature as well as a brief overview of the main features of the smart grid. In summary, the main contributions of this paper are as follows: 
\begin{itemize}
	\item[(i)] a focussed taxonomy of non-cooperative games allowing for a clear classification of the papers under review. We introduce four characteristics that will later allow us to classify the respective games.
	\item[(ii)] a comprehensive brief of the smart grid architecture that underlies the trading activities, with a focus on RE generation and energy storage.
	\item[(iii)] a detailed review of recent advances in `local' energy trading using game--theoretic methods.
\end{itemize}

%\begin{itemize}
%	\item[(i)]an extensive description of a complete GT-based energy trading framework. This includes a taxonomy of GT and an introduction to smart grid architecture focussing on RE generation and energy storage.
%	\item[(ii)]a detailed review of relevant literature about energy trading using game--theoretic methods. 
%\end{itemize}

The paper is structured as follows: Section~\ref{sec:2_gtBasics} focusses on a taxonomy of GT tailored to the concepts that are relevant for the reviewed trading scenarios. Section~\ref{sec:3_etBasics} comprises three parts: a smart grid architecture overview, RE generation, and energy storage. The key contributions of the latest literature is reviewed in Section~\ref{sec:4_gtANDet}. We conclude with a critical evaluation of the current shortcomings.

Please note that in order to increase the outreach of this work, we omit all formulas and instead use only words to present the ideas and concepts behind the game--theoretic approach.

%% file: sec/2_gtBasics.tex
In this section, a brief overview of the vast field of GT is presented, focussing on concepts relevant to energy trading in the smart grid. For more insight into GT, suitable references are provided where appropriate

At the most abstract level, one can classify games into direct and indirect games. 
Direct GT aims to find optimal strategies for players (cf.~\cite{vonNeumann1944}), while indirect GT is concerned with designing games such that certain outcomes will be achieved by rational players (cf.~\cite{Shoham2009}).
As the latter has not played a major role in energy trading to this point in time, we focus our attention on direct games. In particular, non-cooperative games with selfish players, i.e. games in which each player is only concerned about their own outcome. 

\begin{figure}[!t]
	\centering
	\includegraphics[width=0.38\textwidth]{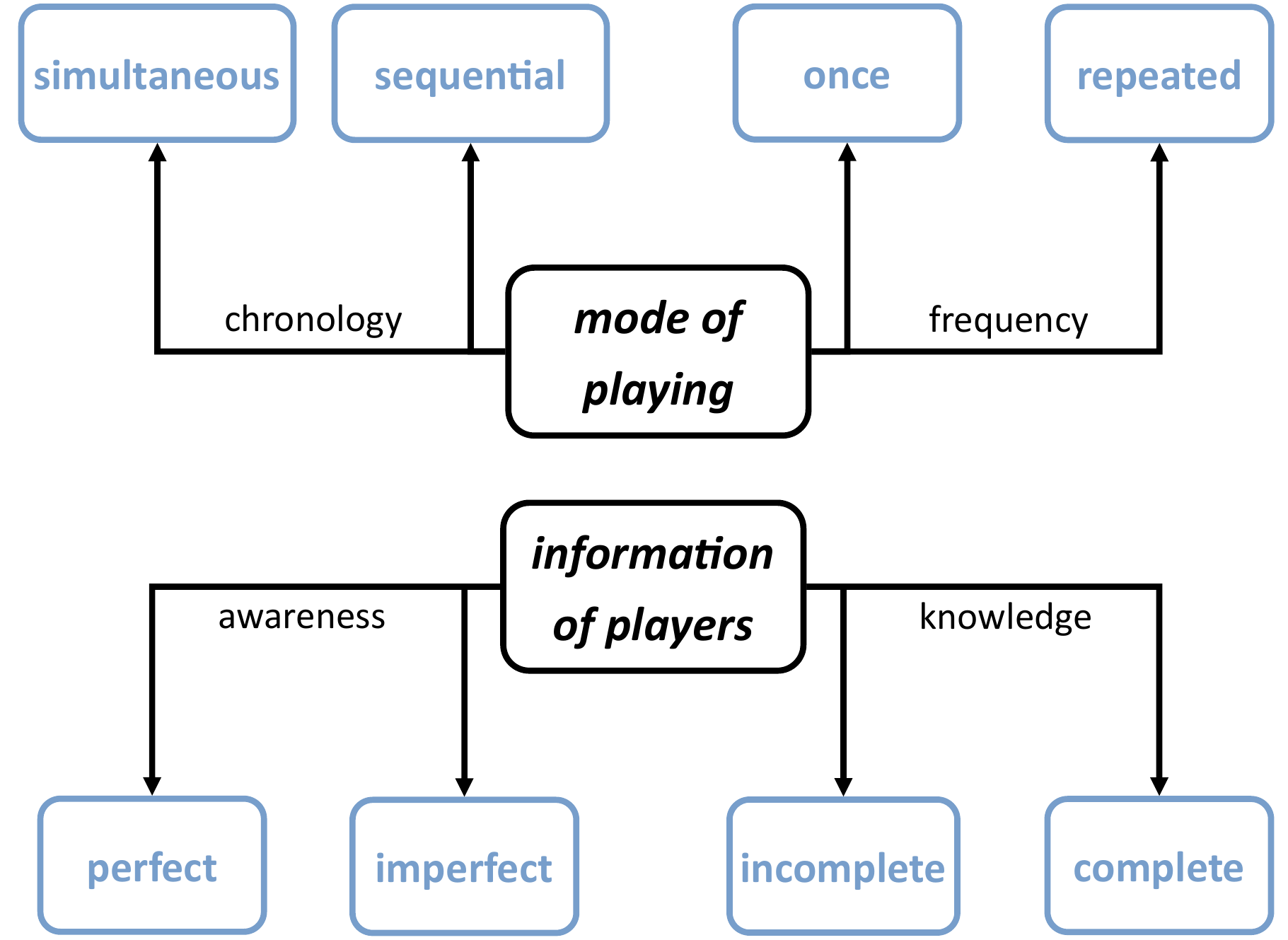}
	\caption{Taxonomy of non-cooperative games split into `mode of playing' and `information of players'.}
	\label{fig:gtTaxonomy}
\end{figure}

To the best of our knowledge, a generally accepted characterisation of games cannot be found currently, as many properties overlap in their classification. In order to introduce a consistent framework, we propose to talk about the `mode of playing' and the `information' each player possesses (cf.~Fig.~\ref{fig:gtTaxonomy}), which in turn leads to four key properties: Frequency, chronology, awareness, and knowledge. There are other criteria, e.g. `value' or symmetry of games~\cite{vonNeumann1944}, but the ones discussed here are sufficient to cover the important aspects for our review. 

(i) \textit{Frequency of play}: Here, we differentiate between games that are played once and games that are played repeatedly. The repetition of a game with the same opponent usually results in different behaviours, as the players have to consider the impact of their actions on the opponent for the next round. 
The utility function, i.e. the outcome for each player, for such games is usually a (weighted) average over the payoffs of each round. The closely related topic of learning in games is discussed in~\cite{Shoham2009}. 

(ii) \textit{Chronology of play}: This refers to either simultaneous or sequential games (cf.~Fig.~\ref{fig:gtTaxonomy}). In a sequential game players move in turns and eventually reach the end of the game where the outcome is defined by a utility function. Moreover, in each turn players might have different actions available. In contrast, players of simultaneous games do not have the ability to react to their opponent. They choose their actions at the same time. This is why they are also called `one-shot' or `static' games. 

%\begin{figure}[!t]
%	\centering
%	\includegraphics[width=0.44\textwidth]{./fig/RPS}
%	\caption{Tree-structure illustration of the Rock-Paper-Scissors game. Arrows represent possible moves: Rock (R), Paper (P), Scissors (S). Each leaf shows the utility function for both players, i.e. (outcome player 1, outcome player 2), with 1 denoting a win, 0 denoting a draw, and -1 denoting a loss. The dotted line shows that player 2 is not aware of which move was played by player 1.}
%	\label{fig:RPS}
%\end{figure}

Note that there is an important difference between a repeated static game and a sequential game. 
Whereas the utility function can be evaluated after each round of a repeated game, it can only be evaluated once at the end of a sequential game. An important class of sequential games are Stackelberg games~\cite{Stackelberg2010}. They originated within an economic application where an established company and a startup compete for market share. The sequential nature is expressed by the burden (or chance) of the bigger company to move first, while the startup can react to the respective decision. More generally, the game exhibits a leader-follower structure. In the energy context we will see a dominant application of this structure, where the seller(s) takes the role of the leader, while the buyer(s) act as followers.

%A way to represent sequential games is in form of a tree structure (cf. Fig.~\ref{fig:RPS}). Each node stands for a certain player and the links originating at the node show their current move set. The utility function is only defined for the leafs of the tree. This makes the analysis of such games more difficult, as no inbetween evaluation is available. 
%The solution of such a game can be obtained by backward induction~\cite{Shoham2009}. The backward induction algorithm defines values for the utility at each node of the tree. The equilibrium is then achieved by best response at every node. Unfortunately, the procedure might be intractable, for instance for chess or go, as the number of possible board configurations is extremely large~\cite{Goldberg1987}.

Von Neumann~\cite{v.Neumann1928} pointed out that one could model a simultaneous game as a sequential game with players being unaware of the other player's move. His idea leads to the next classifier.

(iii) \textit{Awareness of players}: In the literature, one usually refers to perfect and imperfect information. The game of chess serves as a good example of a perfect information game. At every stage of the game, each player knows exactly about the history and in principle (though intractable \cite{Goldberg1987}) about all future moves and their respective outcomes. In an imperfect information game the situation is different. If a player has imperfect information it means they are not aware of the move that has been played before, yet they still know about the general structure of the game, all the utility values, and all possible actions. 
%In the tree representation this unawareness is shown by a dotted line (cf.~Fig.~\ref{fig:RPS}). Player 2 is not aware of which move was played by player 1 and has thus imperfect information of the game.

(iv) \textit{Knowledge of players}: The previous example showed a complete (but imperfect) information game. 
If the knowledge of a player is incomplete, they might not know about the payoffs, strategies, or structure of the game. Such a game is referred to as Bayesian game. The beliefs of a player about the information of their opponents is decoded in their `type' (cf.~\cite{Harsanyi1967, Myerson2004}).

%In a series of papers from 1967-68 \cite{Harsanyi1967, Harsanyi1968, Harsanyi1968a}, Harsanyi gave a first formal definition of such a game, for which he later obtained the Nobel prize. 
%He introduced the notion of the `type' of a player in which all their private information and beliefs are summarised (cf.~\cite{Myerson2004}).
%A key realisation for Harsanyi was that all uncertainties about the game can be captured by uncertainties about the payoffs. Nowadays, such a game is referred to as Bayesian game and mathematically consists of 5 ingredients: (1) a set of players, (2) a set of actions for each player, (3) a set of types for each player, (4) payoff functions for all combinations of types and actions for each player, and (5) a common prior probability function over the types of each player. The last one describes the beliefs of all players regarding another player's type.

%% file: sec/3_etBasics.tex
\begin{figure}[!t]
	\centering
	\includegraphics[width=0.35\textwidth]{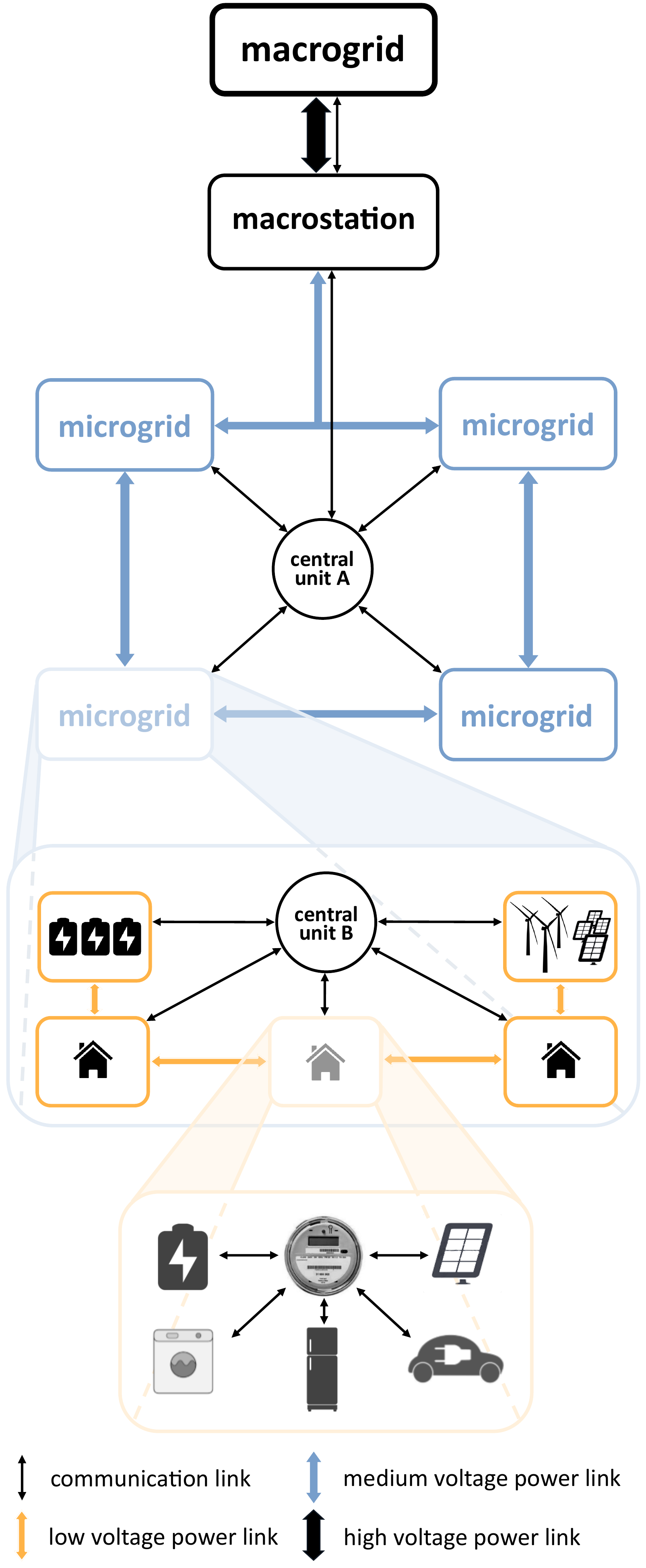}
	\caption{Abstract representation of the envisioned smart grid architecture. Based on decentralisation, two way communication, and energy transfer, a layered structure from the macrogrid on the top to an individual smart home at the bottom is shown.	
	Around \textit{central unit A}, a collection of multiple MGs that are able to exchange energy among each other and communicate through a central unit is shown.
	Zooming into it, the structure within an MG is shown, highlighting the fact that it consists not only of smart homes but also RE generation and medium scale storage facilities. Similar to the layer above it, a central unit manages the direct exchange of energy between the entities.
	Another zoom in shows the smart home which possesses a smart meter allowing it to communicate with its individual RE generation, energy storage system, and various household appliances.}
	\label{fig:smartGridArchitecture}
\end{figure}

A smarter power grid structure is needed to cope with the ever increasing demand while reducing greenhouse gas emissions. Within the current structure, energy is sent from a central large power station to substations. The substations distribute the energy to low voltage subgrids. A failure at a substation could lead to a wide area outage. This is a direct result from the centralised architecture of the grid. The new generation of power grids thus aims for decentralisation. The smart architecture comprises many (almost) self-sustaining MGs that locally produce their own energy. The one-way communication and transmission links are replaced by two-way processes. Key elements for self-sustained operation and a green future are (i) RE generation and (ii) suitable energy storage systems. 

Fig.~\ref{fig:smartGridArchitecture} gives an overview of a smart power grid as envisioned in parts by the authors of the papers under review in Section~\ref{sec:4_gtANDet}. Its inherent decentralised structure is best understood by looking at it from the bottom to the top. The bottom panel is an abstract representation of how a household of the future, i.e. a smart home, might look like. Through its own RE generation and energy storage system, it is able to function independently from the macrogrid. 
All the appliances are continuously monitored by a `smart meter' which also serves as the communication link to the macro/micro grid.

As shown in Fig.~\ref{fig:smartGridArchitecture}, a set of these smart homes together with generation and storage of energy comprises an MG. The notion of MGs is not uniquely defined within the literature. 
A definition from the \textit{US. Department of Energy Microgrid Exchange Group} reads as follows~\cite{USDepartmentofEnergyMicrogridExchangeGroup2016}:
\begin{quote}
A microgrid is a group of interconnected loads and distributed energy resources within clearly defined electrical boundaries that acts as a single controllable entity with respect to the grid. A microgrid can connect and disconnect from the grid to enable it to operate in both grid-connected or island-mode.
\end{quote} 
For privacy reasons, all the communication and organisation of energy transfers is managed by an independent central operator. Nevertheless, smart homes are still able to transfer energy between each other and the storage facilities directly.
In many scenarios we review in Section~\ref{sec:4_gtANDet}, the energy trading algorithms are implemented and executed through the operator.

Similar to within an MG, the communication between multiple MGs is managed by a central unit (cf. Fig.~\ref{fig:smartGridArchitecture}), while energy transfer is possible directly between them. Energy from the utility companies is fed in through the macrostation which connects to the macrogrid via high voltage, long distance power lines.\\

Future generation of RE on the smart home-- and MG--level will rely heavily on wind and photovoltaic (PV) technologies~\cite{Ipakchi2009}. Following their nature, they are usually referred to as variable RE resources. To make them work reliably and efficiently, there are several technical and economical challenges as outlined below.

The variability and uncertainty can be seen in the fact that solar output changes rapidly with the appearance of clouds, or that on average, a wind farm produces 40 per cent of the time of the year, making it very hard to predict the hourly output~\cite{Speer2015}. Another challenge is to balance supply and demand in scarcity and surplus situations. For solar power the highest production rates are achieved during the day which might not coincide with the demand of the respective household. Solutions to these problems are mainly based on better forecasting (e.g.~\cite{Dolara2015,Leva2015}) and demand-side response (e.g.~\cite{Ilic2012,Speer2015}).
%Forecasting techniques based on statistical models and artificial neural networks are used to tackle variability and uncertainty with great success. Vaz\etal~\cite{Vaz2016} showed that by incorporating data from neighbouring PV systems they can forecast PV power production for the next month within small error margins at 15 minute granularity. Similar achievements were obtained by Shayanfar\etal~\cite{Shayanfar2016} for wind power prediction. 

The load-balancing problem is mainly approached from two different sides: Demand-response systems and energy storage. 
The utilisation of an energy storage system allows the user to save surplus energy for times of low production. 
Aside from the technical difficulties, there are also economical challenges that come with the integration of RE generation. Most critical are capital-intensive grid upgrades. For instance, new offshore wind farms require new power lines and additional rooftop PVs might accelerate fatigue of existing components. Solutions range from dynamic line rating~\cite{Speer2015} to better load-forecasting~\cite{Bichpuriya2016} and grid scale energy storage systems. \\
%Wang\etal~\cite{Wang2016} presented a cooperative planning framework for MGs, where they analysed recorded weather data from Hong Kong to establish the best compromise in location and investment allocation of wind and solar farms.\\		
			
Energy storage systems do not only help to balance the operation with diurnal RE generation but also make fast reacting, high-emitting peak power plants obsolete~\cite{Gallo2016}. Moreover, they contribute to an overall improvement of chain efficiency and smoothing of frequency and voltage fluctuations~\cite{Bayram2014}. Eventually, this would result in a more reliable and secure network.
One kind of energy storage system is called electricity energy storage. It converts electricity into another form of energy and then restores electricity back from it. A comprehensive overview over this subclass of storage and others can be found in~\cite{Gallo2016}. 
%A categorisation of electricity energy storage systems is usually done by which immediate energy form is used: mechanical systems store the energy in the form of kinetic or potential energy by pumping, compressing, or accelerating; electrochemical batteries make use of reversible reactions; and electrical energy storage systems exploit electrical fields.
Connecting these storage technologies to the layered grid structure in the context of community based storage, many different types are in use. In contrast, smart home storage is almost completely based on electrochemical energy storage. In particular lithium-ion batteries are widely employed, as they are insensitive to temperature, have long lifetimes, need little maintenance, and can be produced to store enough energy to run a household for one or two days completely independently~\cite{Systems2016}. %Additionally, over their whole life cycle they are also cheaper than the alternative lead-acid batteries~\cite{PowerTechSystems2016a}.\\

%% file: sec/4_gtANDet.tex
In this section, we review the latest achievements of game--theoretic approaches in local energy trading. 
%To this end, we divide the review into two parts based on the specific trading approach. Firstly, there are scenarios in which players of the game \textit{directly} exchange energy with one another. Secondly, we investigate games that incorporate \textit{indirect} energy trading. This usually refers to scenarios where selling and buying is realised through a third party entity.
Table~\ref{tab:game_papers} gives an overview of characteristics of the games employed by the respective research groups, based on our taxonomy in Sec.~\ref{sec:2_gtBasics}.
A comparison of the approaches and results is delivered in the final subsection.

\begin{table} 
	\centering
    \caption{Game--theoretic concepts used in the reviewed papers according to the taxonomy established in Section~\ref{sec:2_gtBasics}}
    \label{tab:game_papers}
    \begin{tabular}{c|cc|cc|cc|cc}
        & \rot{repeated} & \rot{once} & \rot{sequential} & \rot{simultaneous} & \rot{perfect} & \rot{imperfect} & \rot{complete} & \rot{incomplete}\\ \hline
	\rowcolor{fancyBlue!25}\cite{Saad2011}    &	&\OK & & \OK & \OK & & \OK & \rule{0pt}{2.6ex} \\
	
	\cite{Kim2013}     &	&\OK & & \OK & \OK & & \OK &\rule{0pt}{2.6ex}\\
	
	\rowcolor{fancyBlue!25}\cite{Soliman2014}a &	&\OK & & \OK & \OK & & \OK &\rule{0pt}{2.6ex}\\
	
	\cite{Soliman2014}b &	&\OK &\OK &  & \OK & & \OK &\rule{0pt}{2.6ex}\\
	
	\rowcolor{fancyBlue!25}\cite{Tushar2014}  &	&\OK & \OK & & \OK & & \OK &\rule{0pt}{2.6ex}\\
	
	\cite{Lee2015}     &	&\OK &\OK  & & \OK & & \OK & \rule{0pt}{2.6ex}\\
	
	\rowcolor{fancyBlue!25}\cite{Yaagoubi2015}&	&\OK & & \OK & \OK & & \OK &\rule{0pt}{2.6ex}\\
	
	\cite{He2016}     &	&\OK & \OK&  & \OK & & &\OK \rule{0pt}{2.6ex}\\
	
	\rowcolor{fancyBlue!25}\cite{Park2016}    &\OK	& & & \OK & \OK & & \OK &\rule{0pt}{2.6ex}\\
	
	\cite{Rahi2016} &	&\OK & &\OK  & \OK & &  &\OK \rule{0pt}{2.6ex}\\
	
	\rowcolor{fancyBlue!25}\cite{Wang2016a}   &	&\OK &\OK &  & \OK & & \OK & \rule{0pt}{2.6ex}
    \end{tabular}
\end{table}

%In this section, we review the latest achievements of game--theoretic approaches in local energy trading. To this end, we divide the review into two parts based on whether the games incorporate time-dependent trading or not. In the latter, the authors are generally interested in how sellers and buyers interact to make a deal in which all participating parties benefit and are thus incentivised to take part. 
%Time-dependent trading is usually concerned with the question of how sellers and buyers behave under given day-ahead market prices and at what times trading will be beneficial. 

%The final subsection gives a comparison of game-theoretic approaches and results.
%Following the structure of the smart grid architecture shown in Fig.~\ref{fig:smartGridArchitecture}, we divide the review into two parts, one focussing on trading among MGs (\textit{layer A}) and the other focussing on trading within an MG (\textit{layer B}).
%The final subsection gives a comparison of game-theoretic approaches and results.
%\begin{itemize}
%	\item MGs --> lower 'resolution' --> aggregated demands of many households/ large generation
%	\item all papers only consider complete day ahead trading (for the whole 24 hours) --> basically no time dependence in the models
%	
%\end{itemize}

%\subsection{Direct Local Trading}
\subsection{Reviews}

In 2011, Saad\etal~\cite{Saad2011} investigated a futuristic scenario for that time: they considered groups of plug-in hybrid electric vehicles (PHEVs) that are able to sell their stored energy back to the main grid. A group consists of $500$ to $1000$ individual PHEVs with a surplus of energy of more than half their total capacity. One could more generally identify each group as an MG.

Each group acts as a single player of a non-cooperative game and decides on a strategy corresponding to the amount of energy that it is willing to trade. The utility function incorporates the trading and reservation prices, the amount of energy to sell, and a term that summarises costs for discharging the batteries. By means of a double auction\footnote{The authors of \cite{Saad2011} explain how the auction is realised and detail the properties that lead to a strategy-proof mechanism.}, the players are incentivised to truthfully reveal their reservation price.

The game is solved iteratively: starting with an initial strategy of selling all the surplus energy, the sellers take turns replying with their respective best response to the current strategy played by all the other participants. During these steps, the energy company serves as the auctioneer directing the communications (cf. \textit{central unit A} in Fig.~\ref{fig:smartGridArchitecture}). It is shown that the proposed algorithm on average converges to an equilibrium in a reasonable amount of iterations. Compared against a greedy algorithm, in which each seller tries to sell as much energy as possible, the average utility of a group of PHEVs is higher using the game--theoretic approach.\\

%%%%%%%%%%%%%%%%%%%%%%%%%%%%%%%%%%%%%%%%%%%%%%%%%%%%%%%%%%%%%%%%%%%%%%%%%%%%%%%%%%%%%%%
%%%%%%%%%%%%%%%%%%%%%%%%%%%%%%%%%%%%%%%%%%%%%%%%%%%%%%%%%%%%%%%%%%%%%%%%%%%%%%%%%%%%%%%

The idea of using electric vehicles as the future electricity storage units that can also take you from A to B, is considered in the paper by Kim\etal~\cite{Kim2013}. They design a non-cooperative scheduling game for the battery where the users decide between charging the battery, using stored energy for their appliances, or selling stored energy back to the grid. All this is set up in an environment of multiple customers that are connected to an aggregator, which is itself connected to the macrogrid. Participants will declare their expected demand for the following day to the aggregator, allowing it to organise the distribution. The fact that they deal with electric vehicles instead of stationary batteries is modelled within a constraint that denotes times of the day where it can be neither charged nor discharged.

Since the aggregator is interested in making a profit on its own, a tiered billing function is implemented that charges a higher price for heavy users, i.e. users that demand more energy than average. This can also be seen as a measure of fairness, as otherwise these heavy users would drive up the price for all the other users. The need for such a pricing mechanism is justified, because the model is applied to a mixture of residential and industrial customers.

Another consideration to safeguard the aggregator is the incorporation of uncertainty. Whether talking about the uncertainty of weather conditions or the rightfulness of the declared behaviour of the participants, it will eventually lead to uncertainty in the demand. It is assumed that these variations can be bounded by the aggregator based on historical knowledge. Ensuing from the worst-case scenario leads to the analysis of a robust game. Tests show that this increases the social welfare more than twice. More generally, they show that the trading ability improves the social welfare outcome.\\

%%%%%%%%%%%%%%%%%%%%%%%%%%%%%%%%%%%%%%%%%%%%%%%%%%%%%%%%%%%%%%%%%%%%%%%%%%%%%%%%%%%%%%%
%%%%%%%%%%%%%%%%%%%%%%%%%%%%%%%%%%%%%%%%%%%%%%%%%%%%%%%%%%%%%%%%%%%%%%%%%%%%%%%%%%%%%%%

Soliman\etal~\cite{Soliman2014} consider energy storage in the context of scheduling and reducing the peak-to-average-ratio of the demand. They start by designing a smart energy cost function under the conditions that: (i) it is an increasing, strictly convex function, (ii) it pays users to sell energy, and (iii) the price for buying from the utility company is always higher than selling the same amount of energy. These restrictions lead to a stable system with a unique optimum. In order to formulate a scheduling problem, shiftable and non-shiftable appliances are taken as a starting point. Furthermore, it is assumed that the total load of each consumer at a certain time interval of the following day is described by the sum of external power that is bought from the grid, internal power from their own storage device, and an amount that is used to charge their batteries. 

Starting from this initial setup, two different games are proposed. In the first one -- a static non-cooperative game -- the utility sets a cost function that is valid for the following day and the consumers play a `scheduling-game' searching for a strategy that will minimize their respective costs. As users are allowed to sell energy back to the grid, this runs into the phenomenon of a `reverse peak', which happens when users buy extra energy at times of low costs and sell it during peak hours. The second game provides a solution to overcome this problem by making the utility company a participant of the game. It can then adjust the prices in response to the schedules proposed by the consumers. This is a typical leader-follower structure that defines the Stackelberg game (see section \ref{sec:2_gtBasics}). 

A strong result of this paper is the formal proof that the Stackelberg equilibrium is equivalent to the solution that minimises the peak-to-average-ratio. Moreover, their simulations provide evidence that: (i) scheduling with storage always outperforms scheduling without storage in terms of peak-to-average-ratio and cost, (ii) in a scenario with selling back only the total amount of storage matters, and (iii) the consumption profile for the Stackelberg case is almost perfectly flat.\\

%%%%%%%%%%%%%%%%%%%%%%%%%%%%%%%%%%%%%%%%%%%%%%%%%%%%%%%%%%%%%%%%%%%%%%%%%%%%%%%%%%%%%%%
%%%%%%%%%%%%%%%%%%%%%%%%%%%%%%%%%%%%%%%%%%%%%%%%%%%%%%%%%%%%%%%%%%%%%%%%%%%%%%%%%%%%%%%

Tushar\etal~\cite{Tushar2014} look at a situation in which a central power station cannot cope with the high demand at a certain point in time and thus buys the needed energy from what they call energy consumers. These energy consumers are represented by electric vehicles, RE farms, and smart homes, i.e. different grid participants that possess energy storage devices and a communication link to the central power station.
Instead of optimizing each individuals' utility, the authors describe a non-cooperative Stackelberg game that opts to achieve a social optimal solution. With this they assure that each player can benefit from participating in the energy trading, implementing a pricing model where the unit energy price might differ for different energy consumers. The model rewards a higher unit energy price to consumers that can only provide small amounts of surplus energy compared to participants with large surpluses. The authors use an iterative algorithm to minimise the costs for the central power station and simultaneously maximise the sum of the utility functions of the energy consumers. The results show that after 1000 independent simulation runs, the algorithm converges quickly and reliably. 
%The authors show within 1000 independent simulation runs that their iterative algorithm that minimises the costs for the central power station and simultaneously maximises the sum of the utility functions of the energy consumers converges quickly and reliably.
%%%%The results show that the iterative algorithm that minimises the costs for the central power station and simultaneously maximises the sum of the utility functions of the energy consumers converges quickly and reliably in 1000 independent simulation runs. 
Comparisons to a standard feed-in tariff scheme show improvement on average utility per consumer and reduced costs for the power station.\\

%%%%%%%%%%%%%%%%%%%%%%%%%%%%%%%%%%%%%%%%%%%%%%%%%%%%%%%%%%%%%%%%%%%%%%%%%%%%%%%%%%%%%%%
%%%%%%%%%%%%%%%%%%%%%%%%%%%%%%%%%%%%%%%%%%%%%%%%%%%%%%%%%%%%%%%%%%%%%%%%%%%%%%%%%%%%%%%

Lee\etal~\cite{Lee2015} study the trading of energy among MGs, where the MGs do not directly trade with each other but rather try to sell surplus energy to the market or buy required energy from it. This is similar to the architecture described in Section~\ref{sec:3_etBasics}, where the MGs are only able to communicate with a central unit, which serves as a mediator between the MGs, and between the MGs and the macrogrid.
It is assumed that sellers might want to keep parts of their superfluous energy for later time periods, while buyers may buy even more energy than needed, possibly for later trading. 

In the multileader-multifollower Stackelberg game proposed, the sellers act as leaders and the buyers act as followers. The specific utility functions for both groups are set up in a way that achieves a certain level of fairness. This means, the surplus energy offered by the sellers is allocated to all the buyers proportionally to their bids, and the payment from the buyers to the sellers is proportional to their sales volume. Due to the specific definition of the sellers' utility, a convenient simplification of the analysis arises. It turns out that the payoff for each seller depends only on their own strategy and the decision of the buyers. As a result, one only needs to run an optimisation algorithm to maximise the sellers' utility given the buyer's strategies. 

The equilibrium solution for the non-cooperative game among the buyers is given in closed-form and only depends on the selling price and the number of players. A neat and reasonable result for the sum of the utility functions of both the leaders and followers, respectively, is shown. Due to the increasing competition between the buyers, the value monotonically decreases when the number of buyers increases. At the same time, the sum of the utility values for the sellers increases, because more costumers allow them to sell more.\\

%%%%%%%%%%%%%%%%%%%%%%%%%%%%%%%%%%%%%%%%%%%%%%%%%%%%%%%%%%%%%%%%%%%%%%%%%%%%%%%%%%%%%%%
%%%%%%%%%%%%%%%%%%%%%%%%%%%%%%%%%%%%%%%%%%%%%%%%%%%%%%%%%%%%%%%%%%%%%%%%%%%%%%%%%%%%%%%

The focus of \cite{Yaagoubi2015} lies on a more local scenario of energy trading, i.e between individual households. To this end, a neighbourhood of up to 50 users is modelled, dividing the consumers into sellers and buyers. In their scenario the sellers have the freedom to specify the price for which they want to sell surplus energy as long as it is smaller than the energy price from the main grid. The buyers will play a non-cooperative game in which they decide on how much energy to buy from which seller. To point out the local character of this trading, the utility function favours transactions with sellers that are close, i.e. with fewer power line hops between buyer and seller.

Similar to the examples in \cite{Saad2011,Soliman2014,Tushar2014} the solution to the game is achieved by an iterative procedure during which buyers exchange best replies to each others' strategies until nobody wants to deviate any more. On the one hand, the clear advantage is that there is no need for a centralised operator managing the transactions. On the other hand, this method might raise privacy concerns among the users as they need to reveal their information to all the other participants.

For testing the game results, the authors also describe a centralised optimisation model which minimises the total system bill. The comparison between the methods shows that even though the buyers in the game try to minimise their individual energy bill, none of them achieves a lower bill than in the centralised optimisation. Furthermore, it is shown that the iterative algorithm converges fairly fast.\\

%%%%%%%%%%%%%%%%%%%%%%%%%%%%%%%%%%%%%%%%%%%%%%%%%%%%%%%%%%%%%%%%%%%%%%%%%%%%%%%%%%%%%%%
%%%%%%%%%%%%%%%%%%%%%%%%%%%%%%%%%%%%%%%%%%%%%%%%%%%%%%%%%%%%%%%%%%%%%%%%%%%%%%%%%%%%%%%

In \cite{He2016}, another Stackelberg seller-buyer structure among MGs, similar to the ones in \cite{Lee2015, Wang2016a}, is employed. In order to make the model more expressive the authors extend this structure to a Bayesian-Stackelberg game, where each player has private information about his current state. As explained in Section \ref{sec:2_gtBasics}, this knowledge together with beliefs about other players states is summarised in the `type' of the player. Here each player is either one of \textit{normal} or \textit{abnormal} type. The abnormal type represents an emergency state where sellers are less keen to sell energy and value stored energy higher. For buyers this emergency state means, that they are willing to bid more money to guarantee the reception of the requested energy.

In addition to the communication with the market (cf. \cite{Lee2015, Wang2016a}), a second channel of communication is introduced between individual MGs. It models a social network among the them, where a weighting variable is used to express the social relation between the respective entities. A close relationship means that the information communicated is trusted and vice-versa. Mathematically the conditional probability distribution over the state of another player is computed in a two stage process. First, the MG estimates the state based on this players broadcasted messages. Second, it updates this estimate using information it obtained from close `friends' in the network, i.e. their beliefs about the respective player.

Within their model, they are able to simulate the effect of trust between the MGs. In particular they show that only partially trusting the messages of others in the social network, will potentially increase their outcome. As a conclusion they argue that this will improve the power quality, but have yet to show this in future work.\\

%%%%%%%%%%%%%%%%%%%%%%%%%%%%%%%%%%%%%%%%%%%%%%%%%%%%%%%%%%%%%%%%%%%%%%%%%%%%%%%%%%%%%%%
%%%%%%%%%%%%%%%%%%%%%%%%%%%%%%%%%%%%%%%%%%%%%%%%%%%%%%%%%%%%%%%%%%%%%%%%%%%%%%%%%%%%%%%

Park\etal~\cite{Park2016} make use of MG structure shown in Fig.\ref{fig:smartGridArchitecture}. The difference here is that the central unit is not only responsible for the communication but also serves as a gatherer and distributor of the energy that is traded among the MGs. For that reason it is assumed to have a rather large storage capacity, sufficient to store all surplus energy of the MGs that produced more energy than needed in a specified time frame.

Unlike all the other papers reviewed in this section, there is no pricing scheme proposed to pay the sellers. By providing energy to the system the respective MG collects points that increases its contribution value. If it runs into a shortage of energy itself at a later time, a high contribution value will allow it to take a bigger chunk of the energy provided by others at that time. All this is organised by the central distributor, whose goal is to maximise a social welfare function. 
%Park\etal~show that the solution to this problem can be understood as a water-filling problem~\cite{Cover1991}. Each request of energy is represented by a tank of proportional volume and width equal to the contribution value. 

As this distribution mechanism is known to every consumer, the non-cooperative game they play deals with the question of how much energy to request. Directly proportional to this amount and inversely proportional to their individual contribution value, each consumer will be assigned a number in a queue\footnote{Please note that this is a simplification of the actual mechanism. See \cite{Park2016} for the full description of the distribution mechanism.}, i.e.~they have to find a strategy in which they are served early enough while minimising the amount of energy necessary to require from the main grid. If not enough surplus energy is available during the specified time frame, MGs at the end of the queue might not receive any energy through this sharing mechanism. 
%The reason for not simply requesting all the necessary energy to cover the demand is that the tanks are positioned on a base which scales with this amount. The players have to find a strategy in which they are served early enough through the water-filling mechanism while minimising the amount of energy necessary to require from the main grid. 
A nice property of the Nash equilibrium for this case is that even if participants deviate from it, the others will not be influenced negatively. This is shown analytically as well as numerically. Moreover, the run time of the algorithm is short, allowing practical implementation.\\

%\subsection{Indirect Local Trading}

%%%%%%%%%%%%%%%%%%%%%%%%%%%%%%%%%%%%%%%%%%%%%%%%%%%%%%%%%%%%%%%%%%%%%%%%%%%%%%%%%%%%%%%
%%%%%%%%%%%%%%%%%%%%%%%%%%%%%%%%%%%%%%%%%%%%%%%%%%%%%%%%%%%%%%%%%%%%%%%%%%%%%%%%%%%%%%%

Rahi\etal~\cite{Rahi2016} investigate a scenario in which there is a probability of an emergency happening. During such an emergency the utility company buys a predefined amount of energy from the existing MGs at a much higher than the usual price per energy unit. The non-cooperative Bayesian game captures the decisions of the MGs whether to sell their excess energy (from local production) at the current price or store the energy with the possibility of selling it during an emergency with higher profits. Similar to the sellers in \cite{Lee2015, Wang2016a}, the action set of each MG comprises of the proportion of excess energy to be sold. They make use of the Bayesian game structure (cf. Section~\ref{sec:2_gtBasics}) to model the lack of knowledge of an individual MG about the amount of excess energy of other MGs. Thus the \textit{type} of each player is continuous. 

For the scenario with two MGs, Rahi\etal found an analytic solution of the game. There are four equilibria in total, that depend on the beliefs of each player about how much they might be able to sell in case of an emergency. They compare these results with an extended model that changes the utility functions according to prospect theory. This theory is employed to represent typical behaviour in decision making under risk and by this diverts from the assumption of rational players. The solution for this model is obtained through a best--response iteration. It shows deviations of up to ten per cent compared to the first model.\\

%%%%%%%%%%%%%%%%%%%%%%%%%%%%%%%%%%%%%%%%%%%%%%%%%%%%%%%%%%%%%%%%%%%%%%%%%%%%%%%%%%%%%%%
%%%%%%%%%%%%%%%%%%%%%%%%%%%%%%%%%%%%%%%%%%%%%%%%%%%%%%%%%%%%%%%%%%%%%%%%%%%%%%%%%%%%%%%

Similar to the research in \cite{Lee2015}, the architecture shown in \cite{Wang2016a} comprises a number of MGs that are connected with the market/aggregator through which they are enabled to trade excess energy. 
For security reasons, all communications are organised through the central independent operating unit. Viewed from the perspective of any of the MGs, this leads to an incomplete information game, as nobody knows about the strategies and payoffs of the others. More specifically, the authors divide the MGs into sellers and buyers, and design a two stage Stackelberg game in which each of these groups tries to find their best actions by means of a reinforcement learning algorithm. 
The same distribution and paying principles as in \cite{Lee2015} based on proportionality are applied. This means there are two utility functions, one for each of the groups of buyers and sellers. Even without explicit knowledge of the strategies of the other participants, it is shown that the learning algorithm converges to a best reply which is equivalent to the solution of the corresponding optimisation problem for the sellers and buyers, respectively. The tradeoff for the increased privacy is the slower convergence of the iterative scheme. In comparison to the iterative solutions found in \cite{Soliman2014,Tushar2014} it takes approximately 100 times more iterations until convergence. Nevertheless, in a standard test case\cite{IEEEPES}, the algorithm converges to the Nash equilibrium, or at least to the closest best response for scenarios where the Nash equilibrium is not part of the action set.

\subsection{Comparisons}

The literature reviewed in this paper features various similarities and differences. 

Firstly, we want to look at the trading capabilities that are modelled in the different games. Whereas in~\cite{Saad2011, Tushar2014, Rahi2016} the players only consider the possibility to sell energy to the grid, \cite{Kim2013, Soliman2014, Lee2015, Yaagoubi2015, He2016, Park2016, Wang2016a} incorporate two types of players, i.e. sellers and buyers. In many of these cases~\cite{Kim2013, Soliman2014, Lee2015, Park2016} the selling and buying is done in what we call an \textit{indirect} fashion. This refers to models in which energy is not exchanged between the individual participants of the game, but rather through an independent third party/operator. In~\cite{He2016, Wang2016a} a third party is only involved to secure the communication between the trading partners, while the energy is directly transferred between them. The only paper within this review that studies a completely decentralised scenario is~\cite{Yaagoubi2015}.

Secondly, we want to look at more specific game--theoretic differences and similarities. The utility function for each of the respective games is the main feature to consider in this case. Almost all utility functions relate to the monetary cost of energy and hence a pricing component is defined. In all cases considered, higher loads result in higher prices (in agreement with the costs for power generation~\cite{Bayram2014}). In 2013, Kim\etal~\cite{Kim2013} make use of a quadratic relation. Soliman\etal~\cite{Soliman2014} extend this notion by means of a logarithmic expression that behaves in a similar manner to the quadratic relation for positive loads, but also gives more reasonable prices for negative loads, i.e. selling back to the grid. Since then, the tendency has been going towards linear pricing functions as utilised by the authors of \cite{Lee2015, Yaagoubi2015, Wang2016a, He2016}.

Choosing the right utility function can also serve as an opportunity to model the incentives of the players. In addition to the component containing a price function, the utility functions can also include other components to help reflect a more realistic scenario. 
Saad\etal~\cite{Saad2011} incorporate the costs for storing energy. The costs for transmission of energy between trading partners are included in the utility function by Yaagoubi\etal~\cite{Yaagoubi2015}. A combination of these additional costs together with a penalty for insufficient amounts of traded energy is represented in Tushar\etal's utility function~\cite{Tushar2014}. Lee\etal\ and Wang\etal~\cite{Lee2015, Wang2016a, He2016} value the energy that is kept in the seller's own storage. Disparate from all of these utility functions is one given by Park\etal~\cite{Park2016}, where a contribution-based energy allocation algorithm is used such that no pricing function is needed. Here, the utility function depends on the ratio between allocated energy and requested energy.

Given the differences in the expressions for the outcomes of the respective players, different goals are pursued in each of these games. The influence of the number of buyers and sellers on the system costs is investigated in \cite{Lee2015, Yaagoubi2015}. The authors of \cite{Wang2016a, Park2016, Tushar2014} are mainly concerned with validating and testing the algorithm developed in their respective papers. This is sensible as they propose the most blue-sky approaches with `reinforcement learning' and `contribution-based' energy trading. Kim\etal\ and Soliman\etal~\cite{Kim2013, Soliman2014} are concerned with the influence of different types of users. In \cite{Kim2013}, different capabilities of battery storage systems are examined. Additionally, the authors look at the influence of uncertainty and how this leads to a more robust model. Similarly, the question of whether it is beneficial to allow consumers to sell their surplus energy back to the grid or not, is studied in \cite{Soliman2014}.

The results of most of these studies, i.e. the solutions of their respective games, are obtained in one of two ways. While the authors of \cite{Saad2011, Tushar2014, Soliman2014, Yaagoubi2015, He2016} make use the best--response iterative approach (cf.~\cite{Shoham2009}), \cite{Lee2015, Park2016, Rahi2016} derive analytical solutions for their models. Only~\cite{Kim2013, Wang2016a} apply more elaborate numerical algorithms. In particular Kim\etal~\cite{Kim2013} use a steepest descent approach, while Wang\etal~\cite{Wang2016a} employ a `learning automaton'.
On average the games have around 14 players. This number seems rather low, but has to be seen under the light of two facts: (i) small models are sufficient for studies that want to give proof of concepts, and (ii) the complexity of such games has to be suitable for real-time computation in realistic scenarios.

%% file: sec/5_conclusion.tex
In Section~\ref{sec:4_gtANDet}, we reviewed state of the art concepts for energy trading using game--theoretic methods. The results show considerable progress has already been made, however many gaps remain thus giving rise to interesting research questions.

Arguably, the biggest gap stems from the usage of data. In many of the scenarios shown, customers are classified as sellers or buyers, based on whether they have a surplus of energy or not. There are barely any models that combine a high quality analysis of demand data with that of RE generation in the context of energy trading. As already highlighted in Section~\ref{sec:3_etBasics}, it is exactly the uncertain and variable nature of RE that can cause problems and thus they must be considered more rigorously. Bayesian games provide the mathematical framework for incorporating uncertainties. Within the notion of \textit{types} of players, one should be able to model uncertainties in demand as well as generation. However, only little work has been done in this direction at this point in time (see e.g.~\cite{Sola2014}). 
Furthermore, there is a lack of long term, quantitative propositions, opposing the merely one-day ahead analyses in most works on energy trading. With this, one could include seasonal effects of RE generation and household-demand thereby resulting in a more realistic model.